# Fabrication of surface nanoscale axial photonics (SNAP) structures with a femtosecond laser

FANGCHENG SHEN[1,2], XUEWEN SHU[1], LIN ZHANG[2], AND M. SUMETSKY[2,*]

[1]Wuhan National Laboratory for Optoelectronics, Huazhong University of Science and Technology, Wuhan 430074, China
[2]Aston Institute of Photonic Technologies, Aston University, Birmingham B4 7ET, UK
*Corresponding author: m.sumetsky@aston.ac.uk

**Surface nanoscale axial photonics (SNAP) structures are fabricated with a femtosecond laser for the first time. The inscriptions introduced by the laser pressurize the fiber and cause its nanoscale effective radius variation. We demonstrate the subangstrom precise fabrication of individual and coupled SNAP microresonators having the effective radius variation of several nanometers. Our results pave the way to a novel ultraprecise SNAP fabrication technology based on the femtosecond laser inscription.**

Fabrication of microscopic photonic devices and circuits with ultrahigh precision and ultralow loss is of great interest due to their potential applications in optical processing [1-5], quantum computing [6], microwave photonics [7, 8], optical metrology [9], and advanced sensing [10]. However, the outstanding fabrication precision achieved in modern microphotonics, which is currently as small as a few nanometers [2, 5], is still far beyond the precision required for practical applications [3, 11]. Surface nanoscale axial photonics (SNAP) is a new fabrication platform which allows to fabricate photonic structures with an unprecedented sub-angstrom precision and ultralow loss [12-14]. SNAP structures are formed at the surface of an optical fiber with nanoscale effective radius variation (ERV). The performance of these structures is based on whispering gallery modes which circulate around the fiber surface and slowly propagate along the fiber axis. The axial propagation of these modes is so slow that is it can be fully controlled by the nanoscale ERV of the fiber [12]. Usually, light is coupled into the SNAP fiber with a transverse microfiber attached to the fiber surface (Fig. 1). Recently several high performance micro-devices, such as coupled ring resonators [13] and bottle resonator delay lines [14], have been demonstrated based on the SNAP platform.

SNAP structures are usually fabricated by local annealing of the fiber surface using a focused $CO_2$ laser [12-14]. Annealing causes the local relaxation of the residual stress introduced during the fiber drawing process and leads to a nanoscale ERV of the fiber. While a very high subangstrom fabrication precision has been achieved using this method [13, 14], the minimum characteristic length of the introduced ERV is relatively large (~50 μm). SNAP structures can be also fabricated using a UV laser beam exposure of a photosensitive fiber [12, 15]. However, the ERV introduced is typically a few nm only, limited by the available magnitude of photosensitivity. In addition, the requirement of photosensitivity of the fiber restricts the applications of this method.

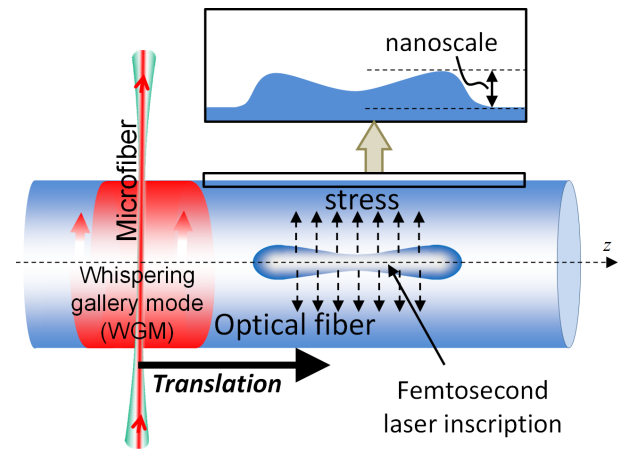

Fig. 1. Illustration of a nanoscale ERV caused by the stress introduced by a femtosecond laser inscription inside the fiber. A microfiber is attached to the fiber surface to excite a whispering gallery mode. The microfiber is also used to characterize the ERV. Inset illustrates a magnified ERV introduced by the femtosecond laser inscription.

Femtosecond laser inscription is a powerful technology that has been successfully employed to fabricate in-fiber devices such as gratings, waveguides, resonators and micro-channels [16-20]. The unique characteristics of a femtosecond laser, such as ultra-short pulse duration, nonlinear nature of the absorption, and extremely high peak powers, enable the three-dimensional fabrication of microscopic objects in various transparent materials [21-23].

In this Letter, we propose and demonstrate the fabrication of SNAP structures using a femtosecond laser. We show that the femtosecond laser inscription introduced along the fiber axis pressurizes the remaining part of the fiber and causes nanoscale

ERV. Using the proposed method, we fabricate individual and coupled SNAP micro-resonators with subangstrom precision.

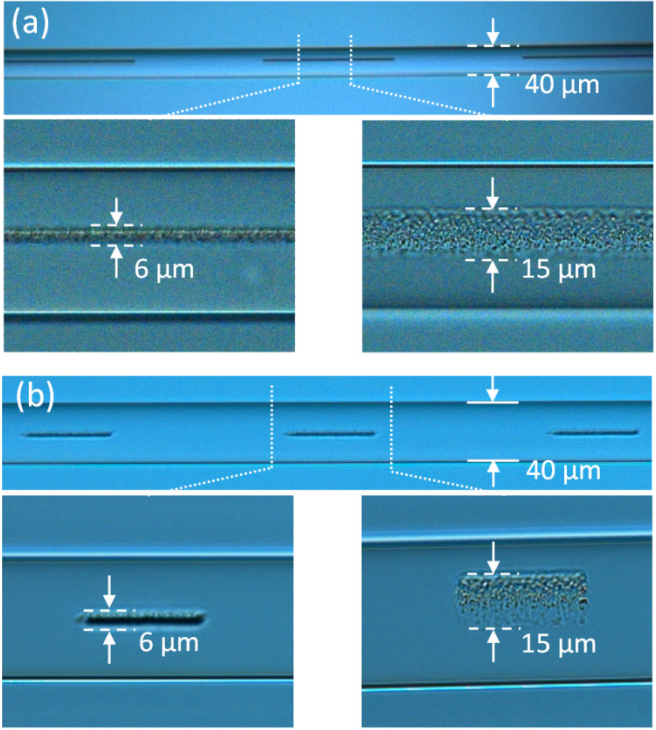

Fig. 2. Microscope images of femtosecond laser inscriptions. (a) – 250 μm long inscriptions separated by 250 μm; (b) – 30 μm long inscriptions separated by 70 μm. The bottoms of (a) and (b) show the magnified pictures of inscriptions taken along the direction of femtosecond laser beam (left) and normal to it (right).

In our experiments, an amplified Ti: Sapphire laser (800 nm wavelength) with a pulse duration of 110 fs and a repetition rate of 1 kHz is used to inscribe the stress rods along the fiber axis. The laser is focused by a 100X microscopic objective (NA = 0.55) into the optical fiber with the radius $r_0 = $ 20 μm. We introduce stress rods of predetermined axial lengths and spacing by switching the laser on and off in the process of translation of the fiber along its axis. Fig. 2 shows the microscope images of rods inscribed periodically with the inscription length of 250 μm separated by 250 μm (Fig. 2(a)) and with the inscription length of 30 μm separated by 70 μm (Fig. 2(b)). The bottoms of Figs. 2(a) and 2(b) show the magnified pictures of inscriptions taken along the direction of femtosecond laser beam (left) and normal to it (right). It is seen that the inscription have strong axial asymmetry being significantly narrower when viewed along the direction of the laser beam (6 μm) than when viewed along the normal direction (15 μm). This asymmetry is caused by the use of a low NA objective, and can be altered using the objectives with other values of NA [22]. Micron-scale irregularities of the introduced inscriptions can be seen in Fig. 2. These irregularities are caused by relatively large pulse energy (~30 nj) and a relatively low repetition rate of the femtosecond laser. In our experiment, the translation speed of the fiber was 20 μm/s for the 250 μm long inscriptions (Fig. 2(a)) and 10 μm/s for the 30 μm long inscriptions (Fig. 2(b)). It is seen from the images of Fig. 2 that the characteristic dimensions of inscriptions do not noticeably change with increasing of the exposure time. We suggest that this happens due to the saturation of the inscription process, which is limited by the repetition rate and power of laser pulses. This effect (to be analyzed elsewhere) causes the saturation of the generated pressure and, consequently, ERV. Remarkably, it took only ~ 10 seconds to introduce a stress rod having a 100 μm length. The inscription speed can be further enhanced with a femtosecond laser having a higher repetition rate.

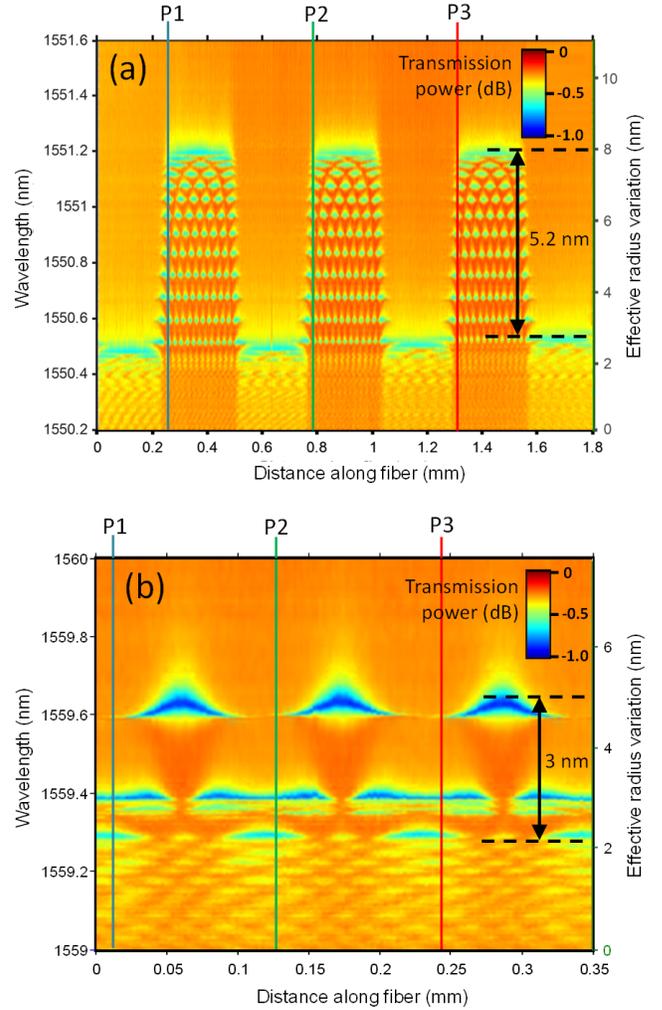

Fig. 3. Surface plots of resonant transmission spectra measured with 2 μm steps along the fiber for (a) 250 μm long inscriptions separated by 250 μm and (b) 30 μm long inscriptions separated by 70 μm. The spectra at positions P1, P2, and P3, which are identically chosen at each microresonator, are compared in Fig. 4.

To characterize the ERV introduced by the femtosecond laser inscription, we use a biconical fiber taper with a microfiber waist [24, 25]. The tapered fiber is connected to the Luna optical spectrum analyzer (OSA) with the resolution of 1.3 pm at 1.55μm. The microfiber waist, which is oriented transversely to the SNAP fiber, is translated along the fiber axis and periodically attached to the SNAP fiber with a 2 μm step (Fig. 1). At each position of the microfiber, the resonant spectrum of excited whispering gallery modes is measured by the OSA.

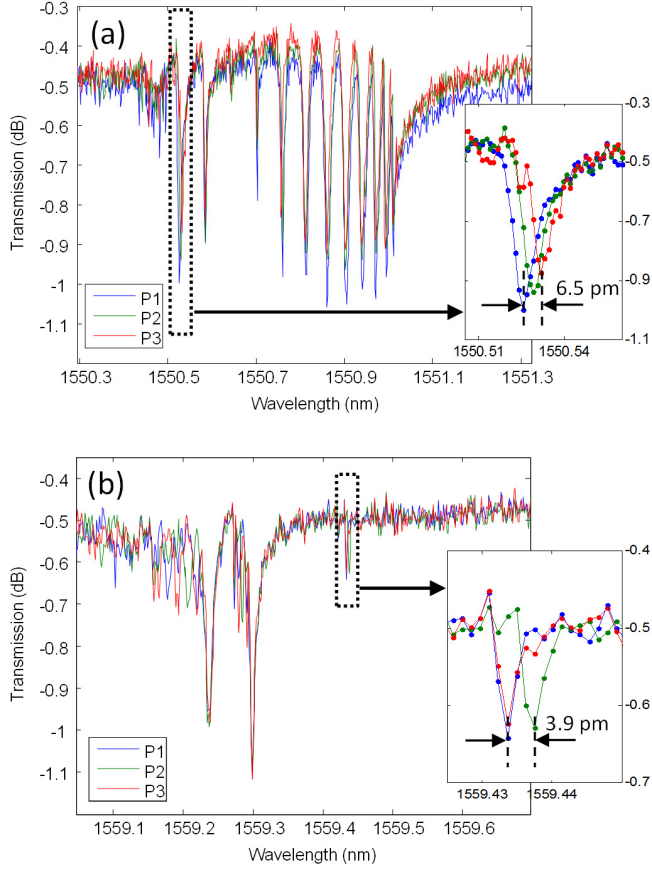

Fig. 4. Testing the fabrication precision by comparison of the spectra of fabricated microresonators. (a) – Spectra of microresonators shown in Fig. 3(a) at positions P1, P2, and P3. Inset compares the resonances with maximum deviation equal to 6.5 pm. (b) – Spectra of coupled microresonators shown in Fig. 3(b) at positions P1, P2, and P3. Inset compares the fundamental axial resonances having the maximum deviation equal to 3.9 pm.

While longer series of microresonators experience stronger mutual deviations, which we attribute to the variation of the power of the femtosecond laser used, we observe a remarkable reproducibility of spectra of a few adjacent microresonators. The surface plots of the measured spectra of three adjacent microresonators corresponding to the inscriptions depicted in Fig. 2 are shown in Fig. 3. This figure demonstrates the similarity of the fabricated microresonators despite the micron-scale irregularities of the introduced inscriptions (Fig. 2), which can be attributed to the averaging of these irregularities in the region outside of the inscription area. The microresonators shown in Fig. 3(a) are created by the 250 μm long inscriptions separated by 250 μm (Fig. 2(a)). It is seen that these microresonators are uncoupled and have 13 resonances (axial eigenvalues) each. The introduced ERV $\Delta r$ is related to the resonant wavelength variation $\Delta \lambda$ by the rescaling relation [12, 24-26]:

$$\Delta r = r_0 \cdot \frac{\Delta \lambda}{\lambda_0}, \quad (1)$$

where the fiber radius $r_0$ = 20 μm, and wavelength $\lambda_0$ = 1550 nm. From this relation, the maximum wavelength variation $\Delta\lambda$ = 0.6 nm in Fig. 3(a) corresponds to the ERV $\Delta r$ = 5 nm. The surface plot of transmission spectra of micro-resonators introduced with 30 μm inscriptions separated by 70 μm is shown in Fig. 3(b). It is seen that these micro-resonators are coupled and have two resonances each. While the fundamental axial resonances are weakly coupled and their splitting is not resolved by the resolution of our OSA, the following lower resonances are strongly coupled (compare with [13]). The axial FWHM dimension of the fundamental mode in these micro-resonators is ~ 30 um, which can be further reduced by increasing the ERV and decreasing the spacing between microresonators, since the modification introduced by femtosecond laser can be highly localized within characteristic dimensions of ~ 1 um [21, 22].

The fabrication precision of the proposed method is determined by comparing the spectra of fabricated resonators. Fig. 4(a) compares the spectra of microresonators shown in Figs. 2(a) and 3(a) at similar positions P1, P2, and P3 of each period indicated in Fig. 3(a). To analyze the precision, we compare the coordinates of resonance minima of these spectra and find that their maximum deviation (shown in the inset of Fig. 4(a)) is 6.5 pm. From the rescaling relation, Eq. (1), this corresponds to the precision of the introduced ERV equal to 0.08 nm. Similarly, Fig. 4(b) compares the minima of the fundamental axial resonances of coupled microresonators shown in Fig. 2(b) and 3(b). The maximum deviation of these minima is 3.9 pm corresponding to 0.05 nm of ERV. Thus, the remarkable subangstrom fabrication precision is demonstrated in both cases.

We analyze the experimental data obtained using a model of thick walled cylinder [27] and approximating the axially asymmetric inscription (Fig. 2) by an axially symmetric cylinder having the effective radius $r_i$. We assume that modifications near the fiber axis in the region $r < r_i$ with characteristic radius $r_i$ pressurize the remaining part of the fiber, $r_i < r < r_0$, and consequently, introduce the ERV. Using the Lame's equation for a thick-walled cylinder [27] we find that the pressure $P_i$, applied to the internal surface of the cylinder $r_i < r < r_0$, is much greater than the atmospheric pressure. Consequently, we find the variation of the fiber radius in the form

$$\Delta r = \frac{2 r_i^2 P_i}{E r_0} \quad (2)$$

where $E$ = 76 GPa is the Young modulus of silica. Since the effect of stress-induced refractive index variation on the optical path length is usually smaller than that of the deformation, the ERV (which combines this radius variation with the stress-induced anisotropic variation of the refractive index) can be estimated from Eq. (2) within a factor of 2. From Eq. (2), the ERV grows linearly with the cross-section area $\pi r_i^2$ of the inscription and is inverse proportional to the fiber radius $r_0$, provided that the pressure $P_i$ introduced by the inscription is independent of its dimensions. We estimate the ERV introduced by the femtosecond laser inscription (Fig. 2) by setting the effective radius of the stress rod $r_i$ ~ 5 μm. Then, for the characteristic $\Delta r$ ~ 5 nm (Fig. 3(a)), the required internal pressure has a value of ~ 0.1 GPa comparable with the characteristic frozen-in stresses in fibers [28]. It follows from Fig. 3(a) and Eq. (2) that a linear axial inscription with a two time greater effective radius, $r_i = 10$ μm, will introduce the ERV which is four times greater than that in Fig. 3(a), i.e., equal to 20 nm. We suggest that the introduced pressure and ERV can be further increased using a femtosecond laser with a higher repetition rate due to the heat accumulation effects [29].

In summary, this is the first demonstration of the fabrication of SNAP structures using a femtosecond laser. Individual and coupled micro-resonators are fabricated with subangstrom precision. The advantage of the femtosecond laser inscription compared to the $CO_2$ laser fabrication of SNAP structures [12-14] is that the former does not rely on the residual stresses of the fiber and, thus, is much more flexible. Furthermore, due to the short pulse duration and much smaller wavelength, a femtosecond laser enables the inscription of ERV with significantly higher axial resolution than a $CO_2$ laser. This is critical for several applications including microscopic delay lines and buffers [1, 14, 30] since it will allow to decrease the characteristic dimensions of SNAP devices and improve their performance. We suggest that similar to the $CO_2$ laser fabrication method [13], a better fabrication precision for longer SNAP structures can be achieved by post-processing. In order to avoid additional losses of microresonators, the inscribed stress regions were separated from the fiber surface by more than 10 μm (Fig. 2). This separation can be reduced down to a few microns without reducing the Q-factor of resonators [31]. The femtosecond laser inscription is irregular at microscale (Fig. 2). However, it leads to the reproducible ERV due to averaging of stress in the bulk of the fiber. It is expected that the performance of the proposed technology can be significantly improved by using a higher frequency femtosecond laser as well as by the optimization of the in-fiber waveguide writing techniques. Overall, the future development of the results of this Letter will lead to a deeper insight into the physics of femtosecond laser inscription, including local stress phenomena, and significant advancement of the SNAP fabrication platform.

**Funding.** Royal Society (WM130110); Horizon 2020 (H2020-EU.1.3.3, 691011); Wuhan National Laboratory for Optoelectronics Director Fund.

**Acknowledgments**. FS acknowledges the China scholarship Council for financial support. MS acknowledges the Royal Society Wolfson Research Merit Award. FS is grateful to Artemiy Dmitriev and Neil Gordon for consultations at the initial stage of this work.